\documentclass[aps,prd,twocolumn,showpacs,superscriptaddress]{revtex4}  
\usepackage{graphicx}  
\usepackage{dcolumn}   
\usepackage{bm}        
\usepackage{amssymb}   

\newcommand{\BABARPubYear}    {07}
\newcommand{\BABARPubNumber}  {012}
\newcommand{\SLACPubNumber} {12393}

\input babarsym

\begin{document}

\begin{flushleft}
\babar-PUB-\BABARPubYear/\BABARPubNumber \\
SLAC-PUB-\SLACPubNumber \\
\end{flushleft}

\title{Measurement of the Relative Branching Fractions of 
{\boldmath $\overline{B} \to D/D^{*}/D^{**} \ell^- \bar{\nu}_{\ell}$} Decays in Events with 
a Fully Reconstructed {\boldmath $B$} Meson}
%
\author{B.~Aubert}
\author{M.~Bona}
\author{D.~Boutigny}
\author{Y.~Karyotakis}
\author{J.~P.~Lees}
\author{V.~Poireau}
\author{X.~Prudent}
\author{V.~Tisserand}
\author{A.~Zghiche}
\affiliation{Laboratoire de Physique des Particules, IN2P3/CNRS et Universit\'e de Savoie, F-74941 Annecy-Le-Vieux, France }
\author{J.~Garra~Tico}
\author{E.~Grauges}
\affiliation{Universitat de Barcelona, Facultat de Fisica, Departament ECM, E-08028 Barcelona, Spain }
% ...
\collaboration{The \babar\ Collaboration}
\noaffiliation
  
\date{\today}

\begin{abstract}
We determine the relative branching fractions 
of semileptonic $B$ decays to charmed final states. The measurement is
performed on the recoil from a fully reconstructed $B$ meson in a sample of 362 million \BB\ pairs collected at the $\Upsilon(4S)$ resonance with the \babar\ detector. A simultaneous fit to a set of discriminating variables is performed on a 
sample of $\overline{B} \rightarrow DX\ell^- \bar{\nu}_{\ell}$ decays 
to determine the contributions from the different channels. We measure $\Gamma(B^- \rightarrow D \ell^- \bar{\nu}_{\ell})/\Gamma (B^- \rightarrow D X \ell^- \bar{\nu}_{\ell})= 0.227 \pm 0.014 \pm 0.016$, $\Gamma(B^- \rightarrow D^{*} \ell^- \bar{\nu}_{\ell})/\Gamma (B^- \rightarrow D X \ell^- \bar{\nu}_{\ell})= 0.582 \pm 0.018 \pm 0.030$ and $\Gamma(B^- \rightarrow D^{**} \ell^- \bar{\nu}_{\ell})/\Gamma (B^- \rightarrow D X \ell^- \bar{\nu}_{\ell})= 0.191 \pm 0.013 \pm 0.019$ for the charged $B$ sample, and $\Gamma(\overline{B^0} \rightarrow D \ell^- \bar{\nu}_{\ell})/\Gamma (\overline{B^0} \rightarrow D X \ell^- \bar{\nu}_{\ell})= 0.215 \pm 0.016 \pm 0.013$, $\Gamma(\overline{B^0} \rightarrow D^{*} \ell^- \bar{\nu}_{\ell})/\Gamma (\overline{B^0} \rightarrow D X \ell^- \bar{\nu}_{\ell})= 0.537 \pm 0.031 \pm 0.036$ and $\Gamma(\overline{B^0} \rightarrow D^{**} \ell^- \bar{\nu}_{\ell})/\Gamma (\overline{B^0} \rightarrow D X \ell^- \bar{\nu}_{\ell})= 0.248 \pm 0.032 \pm 0.030$ for the neutral $B$ sample, where uncertainties are statistical and systematic, respectively.
\end{abstract}

\maketitle 

The determination of exclusive branching fractions of
$\overline{B} \to X_c \ell^- \bar{\nu}_{\ell}$ decays is an essential part of the
$B$-factory program to understand the dynamics of $b$-quark
semileptonic decays and to determine the relevant Cabibbo-Kobayashi-Maskawa matrix elements~\cite{CKM}. The mass of the hadronic 
system $X_c$, recoiling against the leptonic pair, is a crucial 
observable both in the extraction of $|V_{cb}|$, in exclusive semileptonic 
decays, and in isolating $\overline{B} \to X_u \ell^- \bar{\nu}_{\ell}$ decays to determine  
$|V_{ub}|$. It is also needed for the measurement of heavy quark masses and 
other non-perturbative OPE (Operator Product Expansion) parameters from the distribution of spectral 
moments.
This mass spectrum can be better understood by a study of the yields of the different $D$ meson
states in semileptonic decays. Current measurements~\cite{aleph,delphi,babar,belle} show a possible discrepancy
between the sum of exclusive rates and the inclusive semileptonic decay width~\cite{pdg}.
While $\overline{B} \rightarrow D \ell^- \bar{\nu}_{\ell}$ and  $\overline{B} \rightarrow D^* \ell^- \bar{\nu}_{\ell}$ decays account for
about 70\% of this total, the contribution of other states, including resonant and non-resonant $D^{(*)}\pi$ decays, 
is not yet well measured and is a possible explanation of this discrepancy.
 
In this paper, we present a novel technique to extract the exclusive relative branching
fractions for $\overline{B} \to D \ell^- \bar{\nu}_{\ell}$, $\overline{B} \to D^* \ell^- \bar{\nu}_{\ell}$  and
$\overline{B} \to D^{**} \ell^- \bar{\nu}_{\ell}$, with $\ell$ = $e$, $\mu$~\cite{CC}, from an inclusive sample of $\overline{B} \to D X \ell^- \bar{\nu}_{\ell}$ events,  where $X$ can be either nothing or any particle(s) from a semileptonic $B$ decay into a higher mass charm state, or a non-resonant state. We denote by $D^{**}$ any hadronic final
state, containing a charm meson, with total mass above that of the $D^*$
state, thereby including both  $D_J$ excited mesons and $D^{(*)}$+$n \pi$ non-resonant
 states. This technique ensures sensitivity to
all hadronic final states containing a $D$ meson, thus helping us to understand the role of  
 excited $D$ states in saturating the inclusive semileptonic rate. 

This analysis is based on data collected with the \babar\ detector~\cite{detector} at the 
\pep2\ asymmetric-energy $e^+e^-$ storage rings. The data correspond to an integrated luminosity 
of 339.4~fb$^{-1}$ recorded at the $\Upsilon$(4S) 
resonance, or, equivalently, about 362 million \BB\ pairs. A detailed GEANT4-based Monte Carlo (MC) simulation~\cite{Geant} of \BB\ and continuum  $e^+e^- \to f\bar{f}~(f=u,d,s,c,\tau)$ events has been used to study the detector response and its acceptance. The simulation models $\overline{B} \to D^* \ell^- \bar{\nu}_{\ell}$ decays using HQET-based calculations~\cite{HQET}, $\overline{B} \to D \ell^- \bar{\nu}_{\ell}$ and $\overline{B} \to D^{**}(\rightarrow D^{(*)} \pi) \ell^- \bar{\nu}_{\ell}$ decays using the ISGW2 model~\cite{ISGW}, and $\overline{B} \to D^{(*)} \pi \ell^- \bar{\nu}_{\ell}$ decays using the Goity-Roberts model~\cite{Goity}.

We select signal $B$-meson decays in events containing  
a fully reconstructed $B$ meson ($B_{tag}$), which allows us to constrain the kinematics, to reduce the combinatorial background and to determine the charge and flavor of the signal $B$. 
We choose a set of three largely uncorrelated variables to 
discriminate between the different semileptonic decay modes in the reconstructed $\overline{B} \rightarrow D X \ell^- \bar{\nu}_{\ell}$ sample. These are: i) the lepton momentum in the center-of-mass (CM) frame, 
$|\vec{p}_{\ell}|$; ii) the missing mass squared reconstructed with respect to the $D \ell$ 
system, which 
corresponds to the mass of the $X \bar{\nu}_{\ell}$ system, $m_{miss,D}^2 = (p_{\Upsilon} - p_{B_{tag}} - p_D - p_{\ell})^2$, where $p_i$ is the four momentum in the CM frame of the reconstructed state $i$; and iii) the number of 
reconstructed charged tracks in addition to those used for reconstructing the $D \ell$ 
system and the $B_{tag}$, $N_{trks}$. In order to reduce the sensitivity to the modeling of the decays to the different charm states, 
the shapes of these variables are 
extracted from data, using exclusive samples highly enriched in the relevant 
decay modes. 
The relative $D$, $D^*$ and $D^{**}$ contributions are then determined by a 
multiparameter fit to the inclusive sample.
  
We select semileptonic $B$ decays that contain 
one fully reconstructed $D$ meson and that recoil against a fully reconstructed
$B_{tag}$ decaying hadronically.
To obtain a high reconstruction efficiency, the analysis 
exploits the presence of two charmed mesons in the final state: 
one used for the exclusive reconstruction of the $B_{tag}$, and another 
in the semileptonic $B$ decay.

The event reconstruction starts from the semileptonic $B$ decay, selecting a charm meson and a lepton with momentum in the CM frame higher than 0.6 GeV/$c$ and the correct
charge-flavor correlation. Candidate $D^0$ mesons are reconstructed 
in the $K^-\pi^+$, $K^- \pi^+ \pi^0$, $K^- \pi^+ \pi^+ \pi^-$, 
$K^0_S \pi^+ \pi^-$, $K^0_S \pi^+ \pi^- \pi^0$, $K^0_S \pi^0$, $K^+ K^-$, 
$\pi^+ \pi^-$, and $K^0_S K^0_S$ channels, and $D^+$ mesons in the 
$K^- \pi^+ \pi^+$, $K^- \pi^+ \pi^+ \pi^0$, $K^0_S \pi^+$, $K^0_S \pi^+ \pi^0$,
$K^+ K^- \pi^+$, $K^0_S K^+$, and $K^0_S \pi^+ \pi^+ \pi^-$ channels. 
In events with multiple candidates, the candidate with the largest $D$-$\ell$ vertex fit probability is selected.
We then select a fully reconstructed $B_{tag}$ meson 
candidate.
 We reconstruct $B_{tag}$ decays of the type $\overline{B} \rightarrow D Y$, where 
$Y$ represents a collection of hadrons with a total charge of $\pm 1$, composed of $n_1\pi^{\pm}+n_2 K^{\pm}+n_3 K^0_S+n_4\pi^0$, where $n_1+n_2 \leq  5$, $n_3 \leq 2$, and $n_4 \leq 2$. Using $D^0(D^+)$ and $D^{*0}(D^{*+})$ as seeds for $B^-(\overline{B^0})$ decays, we reconstruct about 1000 different decay chains. 

The kinematic consistency of a $B_{tag}$ candidate with a $B$-meson decay is checked using two variables: the beam-energy
substituted mass $m_{ES}=\sqrt{s/4-|\vec{p}_B|^2}$, and the energy difference $\Delta E = E_B -\sqrt{s}/2$. Here $\sqrt{s}$ refers to the total CM  energy, and $|\vec{p}_B|$ and $E_B$ denote the momentum and energy of the $B_{tag}$ candidate in the CM frame. For correctly identified $B_{tag}$ decays, the $m_{ES}$ distribution peaks at the $B$ meson mass, while $\Delta E$ is consistent
with zero.
We select the $B_{tag}$ candidate that has 
no daughter particles in common with the charm meson and 
the lepton from the semileptonic $B$ decay, $m_{ES}$ within the signal region 
defined as 5.27~GeV/$c^2$ $< m_{ES} <$ 5.29~GeV/$c^2$, and the smallest 
$|\Delta E|$ value. Mixing effects in the $\overline{B^0}$ sample are accounted for as described in ~\cite{BBmixing}.

The $\overline{B} \rightarrow D X \ell^- \bar{\nu}_{\ell}$ decays are identified by relatively loose selection criteria. We require the reconstructed  ground-state charm meson invariant mass $M_{D^0}$ ($M_{D^+}$) to be in the range from 1.850 
(1.853)~GeV/$c^2$ to 1.880 (1.883)~ GeV/$c^2$ and the cosine of the angle between the directions of the $D$
candidate and the lepton in the CM frame to be less than zero, to reduce background from non-$B$ semileptonic decays.

After these selection criteria, the sample contains leptons from prompt $B$ decays, as well as cascade $B$ decays, in which the lepton does not come directly from the $B$. There are also background sources of leptons, such as photon conversions and Dalitz $\pi^0$ decays, combinatorial \BB\ background and continuum events, that need to be subtracted. The contamination from cascade $B$ decays, about 15.1~(17.8)\% of the total $B^-(\overline{B^0})$ sample, is subtracted using the simulated MC distributions for these backgrounds.  These events are
reweighted to account for differences among the branching fractions used in our MC simulation and the latest experimental measurements~\cite{thorsten}.
The photon conversion and $\pi^0$ Dalitz decay backgrounds (less than 0.8\% of the
total electron sample) are removed using a dedicated algorithm, which performs the reconstruction of vertices between tracks of opposite charges  whose invariant mass is compatible with a photon conversion or a $\pi^0$ Dalitz decay.
The contributions of combinatorial and continuum $B_{tag}$ backgrounds are estimated from the $m_{ES}$ sideband region 5.21 GeV/$c^{2}$ $< m_{ES} < 5.26$ GeV/$c^{2}$. The $m_{ES}$ distribution is fitted by the sum of a Gaussian function joined to an exponential tail~\cite{CrystalBall} for the signal and an empirical phase-space threshold function~\cite{Argus} for the background. 
Cross-feed effects, i.e. $B^-_{tag} (\overline{B^0}_{tag})$ candidates erroneously reconstructed as a neutral~(charged) $B$,  are corrected using MC  simulations. We estimate the fraction
of cross-feed events in the reconstructed $B^-(\overline{B^0})$ sample to be 6.8\%~(8.1\%).
A total of $6396 \pm 251$~($2981 \pm 122$) events are selected,  
with an estimated purity in $B^-(\overline{B^0}) \to D X \ell^- \bar{\nu}_{\ell}$ of 72\%~(73.8\%). 

Exclusive samples enriched in $D \ell^- \bar{\nu}_{\ell}$, $D^* \ell^- \bar{\nu}_{\ell}$ and $D^{(*)}\pi \ell^- \bar{\nu}_{\ell}$ 
are then selected. Contributions from other semileptonic $B$ decays into charm final states, where one or more particles from a higher mass charm  state are missing (feed-down) or random particles are erroneously associated with the charm candidate (feed-up) are removed. This is done by  selecting signal regions in the missing mass squared distributions $m_{miss,D^{(*,**)}}^2 = (p_{\Upsilon} - p_{B_{tag}} - p_{D^{(*,**)}} - p_{\ell})^2$ corresponding to the exclusive decay being reconstructed.  
The selection criteria are chosen to maximize the sample 
purity. 
We select $D^{*+}$ and $D^{*0}$ candidates by requiring the invariant mass difference between the $D^{*}$ and the $D$ to satisfy the selection criteria in Table \ref{tab:Dstarcuts}. \\ The $B^- \to D^{*0} \ell^- \bar{\nu}_{\ell}$ and $\overline{B^0} \to D^{*+} \ell^- \bar{\nu}_{\ell}$ decays are selected by requiring the missing mass squared
$m_{miss, ~D^{*0}}^2$ to be between $-0.35$~GeV$^2$/$c^4$ 
and 0.5~GeV$^2$/$c^4$ and $|m_{miss, ~D^{*+}}^2|$  to be smaller than 0.55~GeV$^2$/$c^4$, respectively.
Feed-down events from decays to $D^{**}$ states are 
removed by requiring $m_{miss, ~D^{**0}}^2$ and
$m_{miss, ~D^{**+}}^2$ to be incompatible with zero.
The $B^- \to D^{0} \ell^- \bar{\nu}_{\ell}$ and $\overline{B^0} \to D^{+} \ell^- \bar{\nu}_{\ell}$ 
decays are selected by removing feed-down events from 
$D^{*}$ and $D^{**}$ states.
Similar selection criteria are applied for $B^- \to D^{**0} \ell^- \bar{\nu}_{\ell}$ decays, with 
$D^{**0} \to D^{(*)+} \pi^-$, and $\overline{B^0} \to D^{**+} \ell^- \bar{\nu}_{\ell}$ decays, with 
$D^{**+} \to D^{(*)0} \pi^+$.

\begin{table}[!t]
\caption{Invariant mass ranges for $D^{*0}$ and $D^{*+}$ selection.}
\begin{ruledtabular}
\begin{tabular}{cc}
Mode & Selection Criteria \\
$D^{*0} \rightarrow D^0\pi^0$ &  $0.139 < M(D^{*0}) - M(D^0) < 0.145$ GeV/$c^2$ \\
$D^{*0} \rightarrow D^0\gamma$ &  $0.133 < M(D^{*0}) - M(D^0) < 0.151$ GeV/$c^2$ \\
$D^{*+} \rightarrow D^0\pi^+$ &  $0.141 < M(D^{*+}) - M(D^0) < 0.149$ GeV/$c^2$ \\
$D^{*+} \rightarrow D^+\pi^0$ &  $0.138 < M(D^{*+}) - M(D^+) < 0.143$ GeV/$c^2$ \\
\end{tabular}
\end{ruledtabular}
\label{tab:Dstarcuts}
\end{table}

\begin{figure*}[!th]
\includegraphics[scale=0.99]{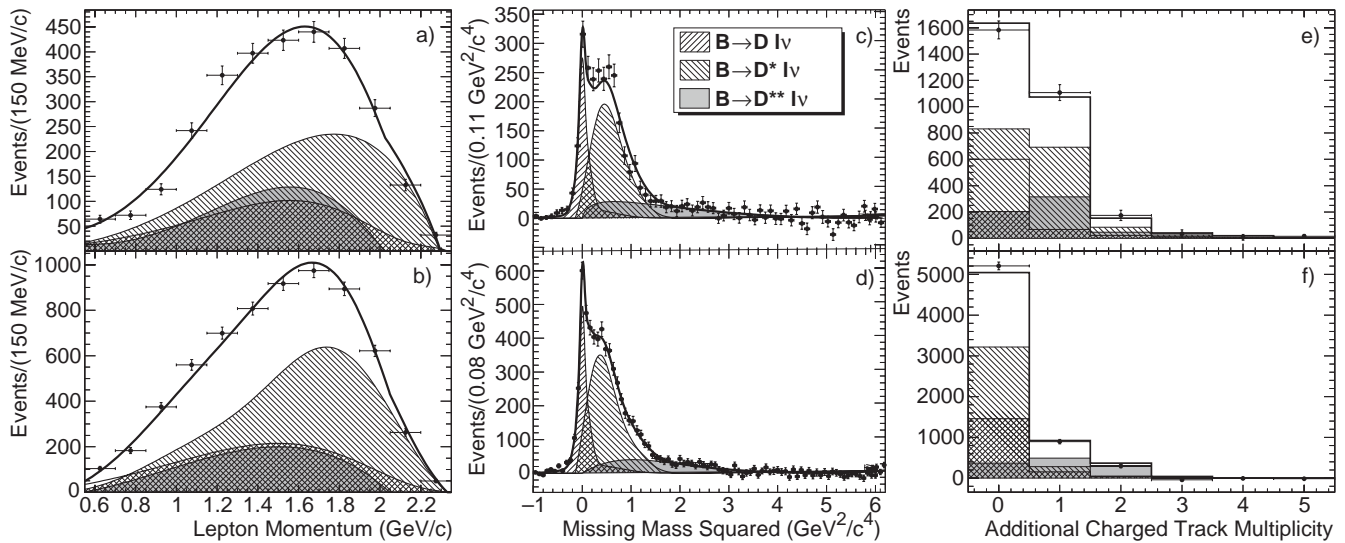}
\label{fig:FitResult}
\caption{ Fitted $|\vec{p}_{\ell}|$ (a,b), $m_{miss,D}^2,$ (c,d), and 
$N_{trks}$ (e,f) distributions for 
$\overline{B^0} \to D X \ell^- \bar{\nu}_{\ell}$ (top) and $B^- \to D X \ell^- \bar{\nu}_{\ell}$ (bottom). The PDFs corresponding to the different exclusive components are superposed with different filling styles.}
\end{figure*}

The probability density functions (PDFs) of the discriminating 
variables, $|\vec{p}_{\ell}|$, $m_{miss,D}^2$ and $N_{trks}$ are determined using the exclusive 
samples. In order to test for possible selection biases in the PDF shapes,
 the inclusive 
distributions for MC samples of $\overline{B} \rightarrow D \ell^- \bar{\nu}_{\ell}$, $D^* \ell^- \bar{\nu}_{\ell}$ and $D^{(*)}\pi \ell^- \bar{\nu}_{\ell}$ events have been compared to those obtained after the exclusive event selection. Good 
agreement is found after accounting for the residual background from feed-down and feed-up
from other modes. The PDFs are parameterized as sums of analytic functions, such as
Gaussians and polynomials, with the exception of $N_{trks}$ which is described using histograms. 

The relative fractions
of $D$, $D^*$ and $D^{**}$ decays in the selected inclusive 
sample of $\overline{B} \to D X \ell^- \bar{\nu}_{\ell}$ events are obtained by a simultaneous $\chi^2$ fit to the inclusive and exclusive  $|\vec{p}_{\ell}|$, $m_{miss,D}^2$ and 
$N_{trks}$ distributions. The relative fractions are floated, constraining their sum to be one, together with the parameters of the functions describing the shapes of
the discriminating variables. 
This results in 
a 35-parameter fit, which ensures that statistical correlations between the different samples are properly 
taken into account and the uncertainties in the exclusive shapes, obtained from samples of
significantly smaller size compared to that of the inclusive sample, are 
correctly propagated into the statistical uncertainties on the $D$, $D^*$ and 
$D^{**}$ relative fractions.
Since this analysis does not reconstruct $D^{**}$ states with neutral pions, 
the $N_{trks}$ distribution for states with the same charged-track multiplicity is used to model these decays: e.g. the 
$B^- \to D^{*0} \ell^- \bar{\nu}_{\ell}$ $N_{trks}$ distribution is used for modeling $D^{**0} (\rightarrow D^{*0}\pi^0) \ell^- \bar{\nu}_{\ell}$ 
decays. For the modes involving a soft charged pion, such as $\overline{B^0}
 \rightarrow D^{*+} \ell^- \bar{\nu}_{\ell}$, the MC prediction for the additional charged-track multiplicity distribution is used to account for  inefficiencies in the reconstruction of the low-momentum particle.    
MC studies show that the PDFs for the $\overline{B} \rightarrow D^{**} \ell^- \bar{\nu}_{\ell}$ component, obtained by the exclusive reconstruction of $\overline{B} \rightarrow D^{(*)}\pi \ell^- \bar{\nu}_{\ell}$ decays, can also be used to parameterize $\overline{B} \rightarrow D^{(*)} n \pi \ell^- \bar{\nu}_{\ell}$ decays in the inclusive $\overline{B} \rightarrow DX \ell^- \bar{\nu}_{\ell}$ sample.  
The fit also accounts 
for feed-down and feed-up decays in the exclusive shapes, fixing the relative contributions  
to the predictions from the simulation. The fit performance has been extensively 
tested using simulated samples with varying fractions of the different decay 
modes. These tests show that the procedure adopted in this analysis is able to 
extract the decay fractions without any significant bias. The statistical
uncertainty obtained by the fit reproduces the scatter of the results from  
independent 
samples, where the bin contents of the distributions have been fluctuated according to their statistical uncertainty.
The fit results for the $\overline{B^0} \to D X \ell^- \bar{\nu}_{\ell}$ and $B^- \to D X \ell^- \bar{\nu}_{\ell}$ distributions of the three variables $|\vec{p}_{\ell}|$, $m_{miss,D}^2$ and 
$N_{trks}$  are shown in Fig.~1. The fit has a $\chi^2$ value of 200 for 212 degrees of freedom for the $B^-$ sample and 204 for 168 degrees of freedom for the $\overline{B^0}$ sample.

\begin{table*}[!ht]
\caption{\label{tab:systB} 
Relative errors (\%) in the determination of $\Gamma(\overline{B} \rightarrow
D^{(*,**)} \ell^- \bar{\nu}_{\ell})/\Gamma(\overline{B} \rightarrow D X \ell^- \bar{\nu}_{\ell})$.}
\begin{ruledtabular}
\begin{tabular}{lccc}
 & $B^-$/$\overline{B^0} \rightarrow D \ell^- \nu_{\ell}$ & $B^-$/$\overline{B^0} \rightarrow D^{*} \ell^- \nu_{\ell}$ & 
$B^-$/$\overline{B^0} \rightarrow D^{**} \ell^- \nu_{\ell}$ \\
\hline
Tracking reconstruction \rule{0pt}{2.6ex} & 3.54/2.36  & 1.3/0.3 & 4.63/3.14\\
Neutral reconstruction & 0.38/0.3 & 0.39/0.31 & 0.41/0.34\\
Lepton identification  & 3.46/3.24  & 3.71/3.57 & 3.51/3.3\\
\hline
$\overline{B} \rightarrow D X \ell^- \bar{\nu}_{\ell}$  Backgrounds \rule{0pt}{2.6ex} & 
0.48/0.78 & 1.69/4.13 & 5.11/7.19\\
$\overline{B} \rightarrow D X \ell^- \bar{\nu}_{\ell}$  Reconstruction efficiency & 
2.35/3.52  & 1.53/2.6 & 3.37/6.43\\
$\overline{B} \rightarrow D X \ell^- \bar{\nu}_{\ell}$ Cross-feed corrections  & 
0.23/0.46  & 0.13/0.56 & 0.71/0.97\\
\hline
$\overline{B} \rightarrow D^{(*,**)} \ell^- \bar{\nu}_{\ell}$ Backgrounds \rule{0pt}{2.6ex} & 
1.81/1.46  & 1.04/1.41 & 1.87/2.26\\
$\overline{B} \rightarrow D^{(*,**)} \ell^- \bar{\nu}_{\ell}$ Feed-down and feed-up corrections & 
1.99/1.49  & 1.31/1.34 & 1.84/2.01\\
$\overline{B} \rightarrow D^{(*,**)} \ell^- \bar{\nu}_{\ell}$ Cross feed corrections  & 
0.74/0.62  & 0.1/0.23 & 1.33/0.74\\
\hline
$|\vec{p}_{\ell}|$ and $m_{miss}^2$ PDFs \rule{0pt}{2.6ex} & 3.27/1.68 & 1.06/1.81 & 1.64/4.8\\
$N_{trks}$ PDF & 0.38/0.9  & 0.91/0.2 & 3.68/0.86\\
$\overline{B} \rightarrow D^{(*)}n\pi \ell^- \bar{\nu}_{\ell}$ & 
0.9/0.73  & 1.1/0.89 & 0.89/0.72\\
\hline
Total Syst.\ \rule{0pt}{2.6ex} & 7.06/6.19 & 5.18/6.71 & 9.88/12.2 \\
\end{tabular}
\end{ruledtabular}
\end{table*}

Several stability checks have been performed. First the sample has been split
into sub-samples based on the lepton flavor and the run period and the fit has been  
repeated for each one of them. Results are consistent within the statistical
uncertainties. As another check, the $\overline{B} \rightarrow D \ell^- \bar{\nu}_{\ell}$ and $\overline{B} \rightarrow D^* \ell^- \bar{\nu}_{\ell}$ branching fractions 
have been determined by a binned likelihood fit to the $m_{miss, ~D}^2$ 
and $m_{miss, ~D^{*}}^2$ distributions respectively, where simulated events are used 
to model the shape of the missing mass squared variables for the $D$, $D^*$ and 
$D^{**}$ exclusive decays and the combinatorial and continuum background. The results are in good agreement with the relative branching fractions  obtained from the fit to the inclusive $\overline{B} \rightarrow D X \ell^- \bar{\nu}_{\ell}$ sample, once we normalize them to the total  semileptonic $B$ branching fraction.

Different sources of systematic uncertainties have been investigated and are given in Table~\ref{tab:systB}.
The first source is due to detector effects, where the size of the uncertainties in the detector response are determined from data control samples.
 Uncertainties related to the reconstruction of charged tracks are determined by evaluating the fit stability using different track selection criteria and by a 
MC study in which we vary the track multiplicity according to the tracking efficiency uncertainty.
The systematic error due to the reconstruction of neutral particles is studied by varying the simulated calorimeter resolution and efficiency.
The systematic uncertainty from lepton identification is estimated by 
varying the tagging efficiency by 2\%~(3\%) for electrons (muons) and the 
misidentification probability by 15\%.

The second main source of systematic uncertainty is related to the selection of the inclusive sample. 
A major contribution is due to background processes, where 
the estimated systematic error is dominated by the uncertainty on the weighting factors used to subtract $B$ cascade decays. The uncertainty 
in the subtraction of the background from the fully reconstructed $B_{tag}$ decays 
is evaluated from the differences 
in the shapes of this background in the sideband and in the signal region using MC predictions. 
The systematic error due to the uncertainty in the amount of flavor cross-feed is computed by varying its fraction  by a conservative 30\%. 
The corresponding systematic uncertainties are evaluated for the exclusive samples.
The analysis, 
relying on decay classification in an inclusive sample, is not sensitive, 
at first order, to reconstruction efficiencies. There remains an uncertainty 
arising from possible differences in efficiencies for the various 
channels, which is estimated from simulation. 

Systematic uncertainties due to the PDFs are estimated by replacing the shapes extracted 
from the exclusive samples with those predicted by our simulation and repeating
the fit. 
Additionally the uncertainty in the relative $D^{**0} \to D^{(*)+} \pi^-$ to 
$D^{**0} \to D^{(*)0} \pi^0$ reconstruction efficiency is accounted for by varying the $N_{trks}$ distribution for the $D^{**0}$ component.

Systematic effects due to $\overline{B} \rightarrow D^{(*)} n \pi \ell^- \bar{\nu}_{\ell}$ events not well parameterized by the  
 $\overline{B} \rightarrow D^{**} \ell^- \bar{\nu}_{\ell}$ PDFs are estimated by repeating the fit with an additional component for these events. The corresponding PDFs are built from a sample of simulated $\overline{B} \rightarrow D^{(*)} \pi\pi \ell^- \bar{\nu}_{\ell}$ events. The  observed difference in the fit results is taken as an additional systematic error. 

\begin{table}[!h]
\caption{Fitted ratios of branching fractions with statistical and systematic uncertainties.}
\begin{ruledtabular}
\begin{tabular}{ccc}
Ratio & $B^-$ (\%) & $\overline{B^0}$ (\%) \\
\hline
{\large{$\frac{\Gamma(\overline{B} \rightarrow D \ell^- \bar{\nu}_{\ell})}{\Gamma (\overline{B} \rightarrow D X \ell^- \bar{\nu}_{\ell})}$}} \rule{0pt}{3.6ex}  &  22.7 $\pm$ 1.4 $\pm$ 1.6 & 21.5 $\pm$ 1.6 $\pm$ 1.3 \\
\hline
{\large{$\frac{\Gamma(\overline{B} \rightarrow D^{*} \ell^- \bar{\nu}_{\ell})}{\Gamma (\overline{B} \rightarrow D X \ell^- \bar{\nu}_{\ell})}$}} \rule{0pt}{3.6ex}  & 58.2 $\pm$ 1.8 $\pm$ 3.0 & 53.7 $\pm$ 3.1 $\pm$ 3.6 \\
\hline
{\large{$\frac{\Gamma(\overline{B} \rightarrow D^{**} \ell^- \bar{\nu}_{\ell})}{\Gamma (\overline{B} \rightarrow D X \ell^- \bar{\nu}_{\ell})}$}} \rule{0pt}{3.6ex}  & 19.1 $\pm$ 1.3 $\pm$ 1.9 & 24.8 $\pm$ 3.2 $\pm$ 3.0 \\
\end{tabular}
\end{ruledtabular}
\label{tab:Results}
\end{table}

In summary, the relative branching fractions for the $B^- \to D^0$, $D^{*0}$, $D^{**0} \ell^-
\bar{\nu}_{\ell}$
and $\overline{B^0} \to D^+$, $D^{*+}$, $D^{**+} \ell^- \bar{\nu}_{\ell}$ decays have
been determined by a multiparameter fit to three discriminating variables in an
inclusive sample of $\overline{B} \to D X \ell^- \bar{\nu}_{\ell}$ events recoiling against a fully
reconstructed $B$ meson. The results are given in Table~\ref{tab:Results}.
Apart from possible isospin violation effects, which are thought to be small, these
three
ratios are expected to be equal for $B^-_u$ and $\overline{B^0_d}$ mesons. The results for
charged and neutral $B$ mesons are compatible within their uncorrelated
uncertainties.
Therefore the relative fractions have been averaged, accounting for correlated
errors.
The results are:
$\Gamma(\overline{B} \rightarrow D \ell^- \bar{\nu}_{\ell})/\Gamma (\overline{B} \rightarrow D X
\ell^- \bar{\nu}_{\ell})$
= 0.221 $\pm$ 0.012~(stat.) $\pm$ 0.006~(uncorr.\ syst.) $\pm$ 0.010~(corr.\ syst.),
$\Gamma(\overline{B} \rightarrow D^{*} \ell^- \bar{\nu}_{\ell})/\Gamma (\overline{B} \rightarrow
D X \ell^- \bar{\nu}_{\ell})$
= 0.572 $\pm$ 0.017~(stat.) $\pm$ 0.016~(uncorr.\ syst.) $\pm$ 0.022~(corr.\ syst.),
$\Gamma(\overline{B} \rightarrow D^{**} \ell^- \bar{\nu}_{\ell})/\Gamma (\overline{B}
\rightarrow D X \ell^- \bar{\nu}_{\ell})$
= 0.197 $\pm$ 0.013~(stat.) $\pm$ 0.013~(uncorr.\ syst.) $\pm$ 0.012~(corr.\ syst.),
where the first uncertainty is
statistical, the second the uncorrelated systematic and the third the
correlated systematic error.
The accuracy of these measurements is comparable to that of the current world
average~\cite{pdg}.

We are grateful for the excellent luminosity and machine conditions
provided by our \pep2\ colleagues, 
and for the substantial dedicated effort from
the computing organizations that support \babar.
The collaborating institutions wish to thank 
SLAC for its support and kind hospitality. 
This work is supported by
DOE
and NSF (USA),
NSERC (Canada),
IHEP (China),
CEA and
CNRS-IN2P3
(France),
BMBF and DFG
(Germany),
INFN (Italy),
FOM (The Netherlands),
NFR (Norway),
MIST (Russia),
MEC (Spain), and
PPARC (United Kingdom). 
Individuals have received support from the
Marie Curie EIF (European Union) and
the A.~P.~Sloan Foundation.


\begin{thebibliography}{99}
\bibitem{CKM}
M.~Kobayashi and T.~Maskawa,
Prog.~Theor.~Phys. {\bf 49},~652 (1973).
\bibitem{aleph}
D.~Skulls {\it et al.}  (ALEPH Collab.),
Z.~Phys.\ C {\bf 73},~601 (1997).
\bibitem{delphi}
P.~Abreu {\it et al.}  (DELPHI Collab.),
~Phys.\ Lett. B {\bf 475},~407 (2000).
\bibitem{babar}
B.~Aubert {\it et al.} (\babar\ Collab.), Phys.\ Rev.~\ D{\bf 71},~051502 (2005).
\bibitem{belle}
D.~Liventsev {\it et al.} (Belle Collab.), Phys.\ Rev.\ D{\bf 72},~051109
(2005).
\bibitem{pdg}
W.~M.~Yao {\it et al.}  (Particle Data Group),
J.~Phys.\ G {\bf 33},~1 (2006).
\bibitem{CC}
Charge conjugate states are always implied unless stated otherwise.
\bibitem{detector}
B.~Aubert {\it et al.} (\babar\ Collab.), Nucl.\ Inst.~\ Meth. A{\bf 479},~1
(2002).
\bibitem{Geant}
S.~Agostinelli {\it et al.}, Nucl.\ Inst.~\ Meth. A{\bf 506},~250 
(2003).
\bibitem{HQET}
J.~Dubosq {\it et al.} (CLEO Collab.), Phys.\ Rev.~\ Lett. {\bf 76},~3898
(1996).
\bibitem{ISGW}
D.~Scora and N.~Isgur, Phys.\ Rev.~\ D{\bf 52},~2783
(1995). See also N.~Isgur {\it et al.},  Phys.\ Rev.~\ D{\bf 39},~799 (1989).
\bibitem{Goity}
L.~Goity and W.~Roberts, Phys.\ Rev.~\ D{\bf 51},~3459
(1995).
\bibitem{BBmixing}
B.~Aubert {\it et al.} (\babar\ Collab.), Phys.\ Rev.~\ D{\bf 69},~111104 (2004).
\bibitem{thorsten}
B.~Aubert {\it et al.} (\babar\ Collab.), Phys.\ Rev.~\ D{\bf 74},~091105 (2006).
\bibitem{CrystalBall}
M.~J.~Oreglia, SLAC-236~(1980);
J.~E.~Gaiser, SLAC-255~(1982);
T.~Skwarnicki, DESY F31-86-02~(1986).
\bibitem{Argus}
H.~Albrecht {\it et al.} (ARGUS Collab.), Z.\ Phys.\ C{\bf 48},~543 (1990).

\end{thebibliography}
\end{document}